# Evidence of a two-step process and pathway dependency in the thermodynamics of poly(diallyldimethylammonium chloride)/poly(sodium acrylate) complexation

L. Vitorazi[1,2], N. Ould-Moussa[1], S. Sekar[3], J. Fresnais[4], W. Loh[2],
J.-P. Chapel[3] and J.-F. Berret[1]*

[1]*Matière et Systèmes Complexes, UMR 7057 CNRS Université Denis Diderot Paris-VII, Bâtiment Condorcet, 10 rue Alice Domon et Léonie Duquet, 75205 Paris, France.*
[2]*Institute of chemistry, Unversidade Estadual de Campinas (UNICAMP), Caixa Postal 6154, Campinas, São Paulo, Brazil.*
[3]*Centre de Recherche Paul Pascal (CRPP), UPR CNRS 8641, Université Bordeaux, 33600 Pessac, France*
[4]*Physicochimie des Electrolytes et Nanosystèmes interfaciaux (PHENIX) UMR 7195 CNRS-UPMC, 4 place Jussieu, 75252 Paris, France*

**Abstract**
Recent studies have pointed the importance of polyelectrolyte assembly in the elaboration of innovative nanomaterials. Beyond their structures, many important questions on the thermodynamics of association remain to be answered. Here, we investigate the complexation between poly(diallyldimethylammonium chloride) (PDADMAC) and poly(sodium acrylate) (PANa) chains using a combination of three techniques: isothermal titration calorimetry (ITC), static and dynamic light scattering and electrophoresis. Upon addition of PDADMAC to PANa or *vice-versa*, the results obtained by the different techniques agree well with each other, and reveal a two-step process. The primary process is the formation of highly charged polyelectrolyte complexes of sizes 100 nm. The secondary process is the transition towards a coacervate phase made of rich and poor polymer droplets. The binding isotherms measured are accounted for using a phenomenological model that provides the thermodynamic parameters for each reaction. Small positive enthalpies and large positive entropies consistent with a counterion release scenario are found throughout this study. Beyond, this work stresses the importance of the underestimated formulation pathway or mixing order in polyelectrolyte complexation.

# I – Introduction

Isothermal titration calorimetry (ITC) is a powerful technique that allows to access the thermodynamic characteristics of physico-chemical reactions in solution.[1,2] The heat released or absorbed during the mixing of interacting components is translated into a set of thermodynamic parameters: the binding constant $K_b$, the binding enthalpy $\Delta H_b$ and the stoichiometry $n$. With the two first quantities, the changes in free energy $\Delta G$ and in entropy $\Delta S$ of the reaction are calculated according to:

$$\Delta G = -RTLnK_b,$$
$$\Delta S = (\Delta H_b - \Delta G)/T \qquad (1)$$

These five quantities are fundamental to understand the type of molecular interactions between the components and to foresee the range of stability or assembly conditions that can be achieved by the reaction. ITC was originally developed in the realm of life science for the study of protein complexes formed by non-covalent interactions.[2,3] ITC was also used to survey the condensation of DNA with proteins or with multivalent counterions. In eukaryotic cells, the DNA/protein biomolecular interactions are fundamental for chromosome compaction, whereas DNA/multivalent counterions are biologically important for non-viral gene delivery. Following Bloomfield and coworkers[4-6] it was shown that the addition of multivalent ions (playing the role of ligands) such as cobalt(III) hexamine and spermidine to DNA double helices led to a two-stage process consisting in *i)* the binding of the DNA strands with the ions and *ii)* their condensation as ligand concentration increased. The two processes were monitored by ITC, and it was shown that both were endothermic, *i.e.* characterized by heat absorption. The binding was associated to a sigmoid-like thermogram of well-defined stoichiometry, binding enthalpy and binding constant, whereas the condensation appeared as a secondary peak at higher concentration. The DNA binding and condensation sequence was also found in assays where cationic surfactants[7,8] or polymers[9-11] were used instead of multivalent counterions.

With the discovery of Layer-*by*-Layer (L*b*L) thin films formed by polyelectrolyte sequential adsorption,[12-14] novel structures of fundamental and applicative interests were designed and the question regarding the thermodynamics of interactions between oppositely charged polymers was raised. In 2006, Laugel *et al.* established a correlation between the growth process of polyelectrolyte multilayers and the heat of complexation between the polyanions and the polycations forming the film.[15] A broad range of polyelectrolytes was investigated, including poly(styrene sulfonate) (PSS) and poly(L-glutamic acid) (PGA) for polyanions, and poly(diallyldimethylammonium chloride) (PDADMAC), poly(allylamine hydrochloride) (PAH) or poly(L-lysine) (PLL) for polycations. It was found that endothermic complexation processes are characteristic of an exponential growth of L*b*L films (e.g. for PGA/PAH and PGA/PLL), whereas a strongly exothermic process corresponds to a linear growth regime of polyanion/polycation pairs (as for PSS/PAH and PSS/PDADMAC[16]). In 2013, Moya and coworkers re-examined this issue and compared the properties of poly(diallyldimethylammonium chloride)/poly(acrylic acid) multilayers in the light of their titration calorimetry responses.[17] These authors found that the binding enthalpy depended on the pH at which the titration was done. From exothermic at acidic pH (pH3) it turned endothermic in alkaline conditions (pH10), an outcome that correlated well with the structures of the L*b*L films. More recently, Priftis *et al.* reported an ITC study using two weak polyelectrolytes, poly(ethylene imine) (PEI) and PGA[18] and observed a complexation thermodynamics driven by entropy and counterion release. ITC showed in this case the presence of two successive processes in the titration curves, one attributed to the ion pairing and the second to the complex coacervation (*i.e.* to the liquid-liquid phase separation between the polymers and the solvent).[18,19] Interestingly enough, the shape of binding curves were similar to those found with oligonucleotides and multivalent counterions.[4,6-9,11,20]

Despite a large number of recent ITC studies on non-biological systems (see also[21-31]), many questions on the thermodynamics of complexation remain unanswered. The first question concerns the microscopic processes involved in polyelectrolyte titrations. It is generally assumed that the measured enthalpy relates to the interaction of pairs of opposite charges belonging to different chains.[18] However, to our knowledge, no systematic study of the structures or phases formed during titration was reported for polyelectrolytes. A number of investigations provided turbidity,[8,18,21,22,31,32] light scattering[8,28,31,33-38] and zeta potential[24,36] measurements for dispersions of oppositely charged species, but not in the exact conditions of titration calorimetry. This approach is still lacking for polyelectrolytes (see however the recent work by Huang and Lapitsky on chitosan and counterions[39]). In terms of interpretation, it is also generally assumed that the existing theoretical frameworks (typically those based on Langmuir adsorption principle and equilibrium reactions[3,24,40]) are valid for polyelectrolyte complexation, although there are by now strong evidences that polyelectrolyte complexes are stable and out-of equilibrium structures.[15,37,38] Another interesting issue concerns the possibility of having a generic ITC behavior for oppositely charged species, as we have seen with oligonucleotides and with some polymer systems.

In the present study, we investigate the complexation between poly(diallyldimethylammonium chloride) and poly(sodium acrylate) (PANa) chains using a combination of three techniques: ITC, static and dynamic light scattering and electrophoresis. To establish a correlation between structural and thermodynamic changes, the same mixing protocols were used. The reason for studying this polyelectrolyte pair in particular stems from our recent work on electrostatic assembly using PANa coated nanoparticles and polycations (e.g. PDADMAC, PEI), which provided remarkably stable nanostructures from opposite charge interactions.[41] Here, we confirm the presence of two sequential processes in the titration calorimetry and identify the associated structures. A phenomenological model based on a Multiple Non-Interacting Sites mechanism[3,24,40] accounts well for the measured binding enthalpies.

# 2 – Materials and Methods
## 2.1 - Materials

Poly(sodium acrylate) (PANa, $M_W$ = 2100 g mol$^{-1}$), poly(acrylic acid) (PAA, $M_W$ = 100000 g mol$^{-1}$) and poly(diallyldimethylammonium chloride) (PDADMAC, $M_W$ = 100000 g mol$^{-1}$) were purchased from Sigma-Aldrich and used without further purification. Molecular structures of PANa and PDADMAC polymers are shown in Fig. 1a. Poly(acrylic acid) is a weak acid and its degree of ionization is controlled by the $pH$. The number of charged monomers was determined from acid-base titration (Supplementary Information S1). The experiments were performed at $pH$ 10 to ensure a full ionization of the chain. PDADMAC in contrast is a strong polyelectrolyte and its degree of dissociation is $pH$ insensitive. The number of positive charges was calculated from the value of the molecular weight of the diallyldimethylammonium chloride monomer (162 g mol$^{-1}$). For electrostatic complexation, we introduce the charge ratio, noted $Z_{+/-} = \frac{[+]}{[-]}$ or $Z_{-/+} = \frac{[-]}{[+]}$, were [+] and [−] denotes the molar concentrations of positive and negative charges, respectively. Depending on the mixing order, $Z_{+/-}$ or $Z_{-/+}$ will be used alternatively. A molar concentration of charges of 20 mM for PDADMAC and for PANa$_{2K}$ corresponds to weight concentrations of 0.32 and 0.19 wt. % respectively.

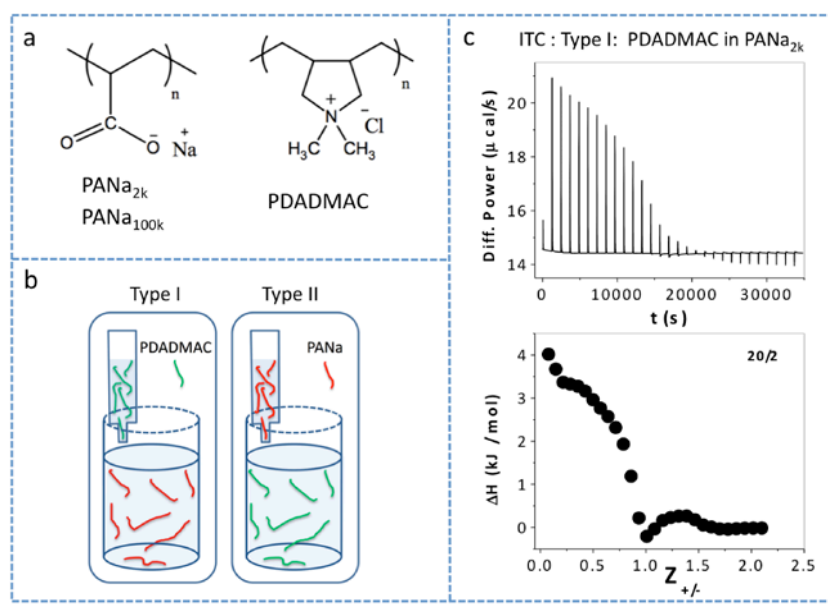

*Figure 1:* a) Structure of polymers used in this study: PANa denotes poly(sodium acrylate) and PDADMAC poly(diallyldimethylammonium chloride). b) Representative titration schemes for Type I and Type II mixing. c) Example of thermogram (upper panel) and enthalpy (lower panel) curves obtained by titrating PANa$_{2K}$ by PDADMAC (Type I).

2.3 - Mixing protocols

Polymer solutions were mixed according to two different protocols: direct mixing and titration. With direct mixing, appropriate volumes of stock solutions prepared at the same charge concentration (20 mM) and same pH ($pH$10) were added at once to solutions of oppositely charged species. The volumes were adjusted to cover charge ratios $Z_{+/-}$ or $Z_{-/+}$ between $10^{-2}$ and 100. The samples were let to equilibrate for 1 day, and their phase behavior was determined by visual inspection and later by dynamic light scattering. Titration consists in a step-by-step addition of a few microliters of a solution into another, each injections being separated to the next by 6 min. These conditions correspond to those of titration calorimetry. To investigate the role of the mixing order, positively charged PDADMAC are mixed to negatively charged PANa$_{2K}$ and *vice versa*. In the following, Type I mixing or titration corresponds to the addition of the positive chains into negative ones, and Type II mixing or titration denotes the reverse.[24,32] A cartoon illustrating the Type I and Type II mixing is shown in Fig. 1b.

2.4 Isothermal titration calorimetry

Isothermal titration calorimetry (ITC) was performed using a Microcal VP-ITC calorimeter (Northampton, MA) with cell of 1.464 mL, working at 25° C and agitation speed of 307 rpm. The syringe and the measuring cell were filled with degased solutions of PDADMAC, PANa at *pH*10. Water was also degased and filled the reference cell.[2,3] Typical charge concentrations were 10 mM in the syringe and 1 mM in the measuring chamber. The titration consisted in a preliminary injection of 2 μL, followed by 28 injections of 10 μL at 10 to 20 min intervals. A typical ITC experiment including the thermogram (differential power provided by the calorimeter to keep the temperature of cell and reference identical) and binding isotherm is shown in Fig. 1c. There, a 20 mM PDADMAC solution is injected into a 2 mM PANa solution at pH 10. Control experiments were carried out to determine the enthalpies associated to dilution. These behaviors were later subtracted to obtain the neat heat of binding.

2.5 - Light scattering and electrophoretic mobility

Light scattering and zeta potential measurements were carried out using a NanoZS Zetasizer (Malvern Instruments). The automatic titrator (Malvern Instruments, MPT-2) coupled to the light scattering and electrophoresis experiments was used to reproduce the same time sequence as that of the titration calorimeter. 10 μL injections were monitored every 6 minutes in Type I and II modes. In the light scattering experiment (detection angle at 173°), the hydrodynamic diameter $D_H$ and the scattered intensity $I_S$ were measured as a function of the time. The time axis was later translated into charge ratios. Both $D_H$ and $I_S$ are indicative of the state of aggregation of the polymers. Complementary experiments performed in the phase analysis light scattering mode (detection angle at 16°) allowed to determine the electrophoretic mobility $\mu_E$ of polymers or of polyeletrolyte complexes, from which the zeta potential $\zeta$ was derived. In this study, the concentration, pH (*pH* 10) and temperature conditions (T = 25 °C) were the same as with the VP-ITC.

2.6 – Optical microscopy

Bright field and phase-contrast images of coacervate phases were acquired on an IX73 inverted microscope (Olympus) equipped with 40× and 60× objectives. Data acquisition and treatment were monitored with a Photometrics Cascade camera (Roper Scientific) and treated with Metaview (Universal Imaging Inc.) and ImageJ softwares. Image treatment was used to estimate the size and dispersion of the coacervate droplets.

# 3 – Results and Discussion

## 3.1 – Phase behavior

Fig. 2 displays images of dispersions prepared by Type I direct mixing. The vials were prepared at charge ratios $Z_{-/+}$ between 0 and 10. Visual observations of the mixed dispersions reveal a maximum of turbidity around $Z_{-/+} = 1$, indicating the formation of inter-polymer structures. Examined by light scattering, the solutions apart from the stoichiometry revealed stable polymer aggregates, commonly known as polyelectrolyte complexes or PECs.[16,33,37,38,42-45] The turbid phase was investigated by phase contrast optical microscopy.

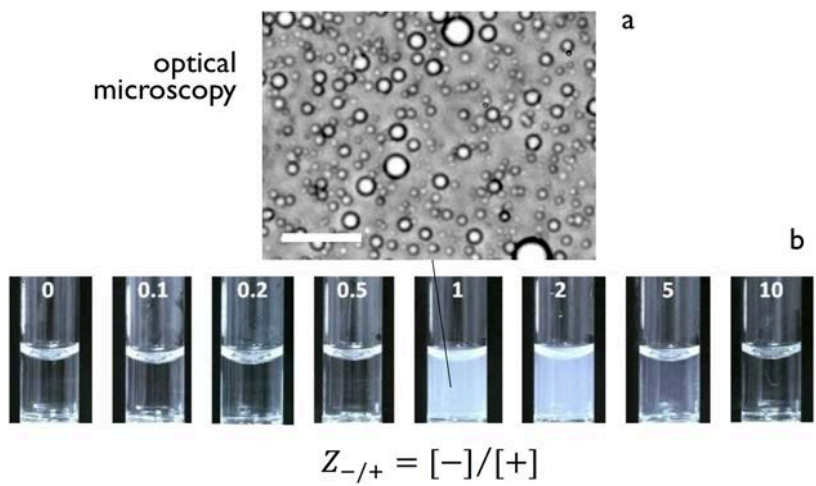

*Figure 2: a) Optical microscopy image of biphasic state obtained by mixing oppositely charged PDADMAC and PANa$_{2K}$ at the charge stoichiometry ($Z_{-/+} = [-]/[+] = 1$) and concentration 20 mM (the bar is 20 μm). b) Images of vials containing the polymers at different charge ratios $Z_{-/+}$ between 0 and 10.*

Fig. 2a shows the presence of stable micron-sized droplets dispersed in an outer liquid phase. The droplet size distribution is independent on the mixing order. The images are indicative of a liquid-liquid phase separation characteristic of a coacervate phase.[18,39,42,46-48] The liquid droplets are made of a concentrated phase of PDADMAC and PANa$_{2K}$ polymers forming polyelectrolyte complexes. The centrifugation or the sonication of the mixed dispersions induces the coalescence of the droplets and the formation of two well-separated phases of different viscosities.[39,46] These data are similar to those found in other oppositely charged polymers systems, including poly(acrylic acid)/poly(N; N-dimethylaminoethyl methacrylate)[46] or poly(ethyleneimine)/poly(aspartic acid).[18] For PDADMAC and PANa$_{2K}$ or PANa$_{100K}$, no solid precipitate was found. Addition of salt (NH$_4$Cl) above 0.6 M dissolves the coacervate phase, suggesting that the entropy driven phase separation occurs under pure electrostatic interaction with no secondary forces at play (from e.g. H-bonding or hydrophobic interaction).[42]

3.2 – Isothermal Titration Calorimetry

Figs. 3a and 3b display the binding isotherms obtained for poly(sodium acrylate) and poly(diallyldimethylammonium chloride) using the Type I and Type II titration respectively. Experiments were performed at concentrations 10/1, 20/2 and 30/3 mM. For Type I experiments (Fig. 3a), the enthalpy curves exhibit a sigmoidal decrease with increasing charge ratio $Z_{+/-}$. Beyond charge neutralization, heat exchanges close to zero indicate the absence of thermodynamic interactions between the added polycations and the phases and structures formed. Throughout the process, the enthalpy is positive and associated with an endothermic reaction. For Type II experiments (Fig. 3b), a similar endothermic behavior is observed, the enthalpy decreasing down to zero at ratios $Z_{-/+}$ around 2. The relatively good superimposition of the thermograms upon concentration change is important because it proves that the thermodynamic processes involved in the titration have the same origin. A careful observation of the data of Fig. 3 reveals that the titration occurs in two steps. The primary process is initiated at low charge ratios and is related to the overall sigmoid-like decrease of the enthalpy. It is followed by a secondary process that manifests itself as a smooth endothermic peak around $Z_{-/+} = 1$ (Fig. 3b). In Fig. 3a, this secondary peak is indeed slightly negative and accounts for the rapid decrease of the enthalpy above 0.5.

Fig. 3c and 3d compare the binding enthalpies at pH7 and 10. Type I and II isotherms at pH7 show typically the same overall behavior as at pH10, the secondary process being slightly less prominent. The enthalpies are also shifted to lower $Z_{+/-}$ (higher $Z_{-/+}$) for Type I (Type II) experiments, respectively. Also observed by turbidity measurements, these shifts are in agreement with the decrease of anionic charges accessible to DADMAC monomers (at pH7, 70% of the acrylate monomers are charged, see Supporting Information S1 and S2). There is another main difference between the two pHs: at pH7, the potentiometric curves for the incremental addition of PDADMAC in PANa$_{2K}$ or *vice versa* are accompanied by a marked proton release and an acidification of the bulk solution. In Fig. S3, it is shown that pH is decreased from 7 to 5 with increasing charge ratio. This result is in agreement with an earlier report on PDADMAC/PANa titrations.[43] Such acidification does not show up in experiments at pH10. The binding enthalpy measured at pH7 could then have two origins: i) the complexation between the two polymers, as at pH10 and ii) the acidification of the PANa chains. At this point, it is difficult to assign which contribution is predominant. Fig. 3e and 3f show the binding enthalpies at concentrations 20/2 in Type I and II conditions (pH10) using 100000 g mol$^{-1}$ molecular weight poly(sodium acrylate). With PANa$_{100K}$, the isotherms are again similar to those found with PANa$_{2K}$, the secondary peak (negative in Fig. 3e or positive in Fig. 3f) being now more pronounced and well separated from the primary process.

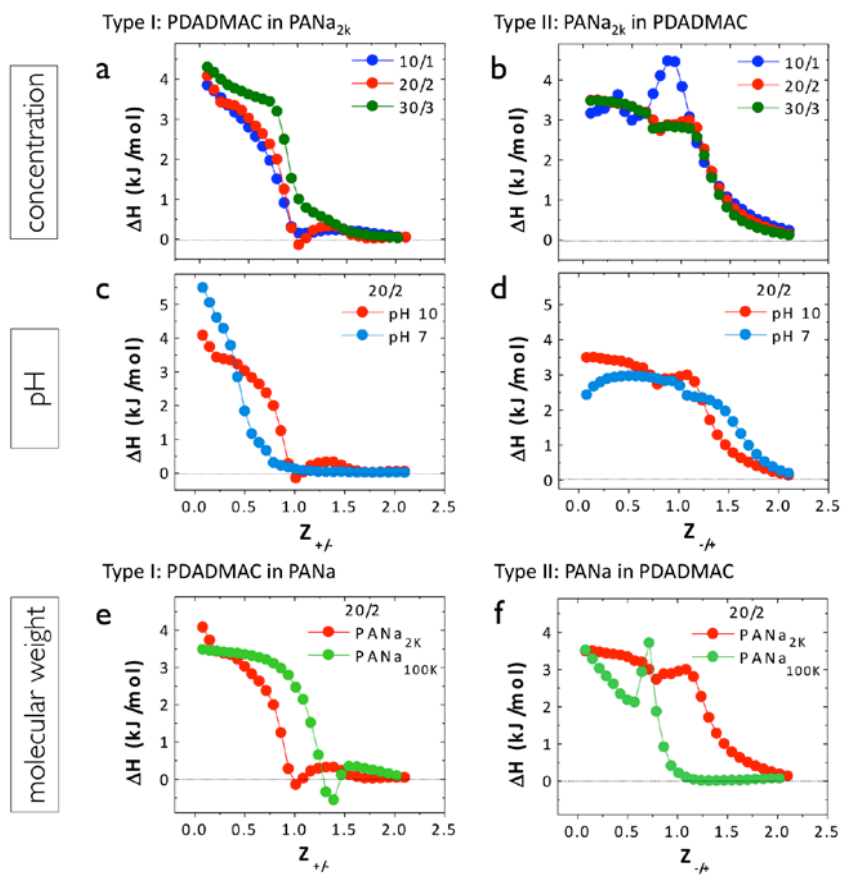

*Figure 3: Binding isotherms obtained in Type I (PDADMAC in PANa) and Type II (PANa in PDADMAC) titration modes at different concentrations (a,b), pHs (c,d) and molecular weight of the PANa (e,f). In all experiments, the temperature was set at T = 25 °C, and when not specified the pH was 10. In all three examples, the enthalpies of binding are positive, indicating endothermic reactions.*

Figs. 4a and 4b illustrates the kinetics associated with the primary and secondary processes for Type I and Type II experiments. In the insets, the differential power is plotted as a function of the time for 29 injections at concentrations 30/3. The main frame of Fig. 4a illustrates the normalized heat at injection numbers 3 ($Z_{+/-} = 0.15$) and 18 ($Z_{+/-} = 1.24$). At the third injection, the differential power represents the intrinsic response function of the calorimeter.[24] The return towards equilibrium then occurs within 100 s. For injections around $Z_{+/-} = 1$ in contrast, a slower kinetics is observed. The slow relaxations around charge neutrality are related to the secondary exo- or endothermic peaks, and have characteristic times of the order of 300 s. Type II titration assays show typically the same behavior (Fig. 4b). The features presented here were observed in earlier work, in particular in studies of polymers and surfactants[8,24] and on multivalent counterions and DNA.[9,10] Interestingly enough, these slow relaxations were not discussed nor related to the formation of coacervate phases. In conclusion to this part, the results obtained by changing the concentration, the pH of the stock solutions and the molecular weight of the polyanions agree well with each other, and confirm the sequence of two titration processes.

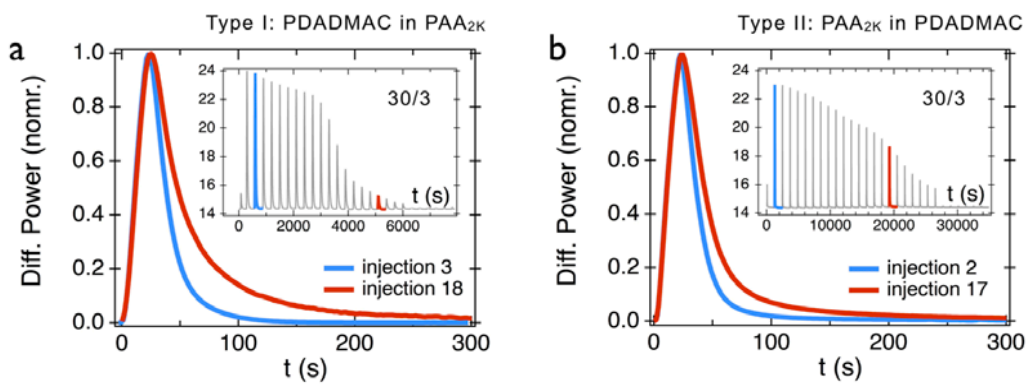

*Figure 4:* a) Normalized ITC differential power versus time for injections number 3 (blue, $Z_{+/-} = 0.15$) and 18 (red, $Z_{+/-} = 1.24$) as observed in the titration of PANa$_{2K}$ by PDADMAC at concentrations 30/3 mM (Type I). b) Same as in Fig. 4a for injections number 2 (blue, $Z_{-/+} = 0.08$) and 17 (red, $Z_{-/+} = 1.16$) as observed in the titration of PDADMAC by PANa$_{2K}$ (Type II). Insets: titration associated thermograms for the two experiments. Longer transient responses of the titration calorimeter correspond to the formation of the PDADMAC/PANa$_{2K}$ coacervate phase.

The data in Fig. 3b display some similarities with those obtained by Alonso et al.[17] on PDADMAC/PANa with slightly different molecular weights. At pH10, both measurements exhibit endothermic responses, but the binding enthalpies measured by Alonso et al.'s are lower than in our case by a factor 6 to 8, probably due to the use of a salted buffer (0.5 M NaCl) and to the screening of the electrostatic interactions. It could be argued furthermore than Alonso *et al.* performed a type II experiment at concentrations (14 mM into 1.4 mM) but did not observe any peak in their thermograms, showing again that salt, macromolecules concentration and mixing order do matter in an ITC experiment.

3.3 – Relating structure to thermodynamic titration
Figs. 5 and 6 display the enthalpy of binding ΔH, the scattering intensity $I_S$, the hydrodynamic diameter $D_H$ and the zeta potential ζ as a function of the mixing ratio for Type I and II titration, respectively. The light scattering and electrokinetics experiments were performed in the same conditions of concentration (20/2), pH (pH10), temperature (T = 25 °C) and injected amounts as those of ITC. For the $D_H$-data, only the hydrodynamic sizes corresponding to the fastest mode of the autocorrelation function were pointed out. Fig. 5a compares the ITC data of Fig. 3a to those obtained by static (Fig. 5b) and dynamic (Fig. 5c) light scattering. Upon addition of polycations, the scattered intensity $I_S$ increases progressively, indicating the formation of PDADMAC/PANa$_{2K}$ mixed structures. The hydrodynamic diameter exhibits a plateau at $D_H = 130 \pm 10$ nm, up to the critical ratio 0.90. Beyond this limit, the diameter increases rapidly and reaches micron-sized values. This size is associated with the formation of the coacervate phase (Fig. 2).

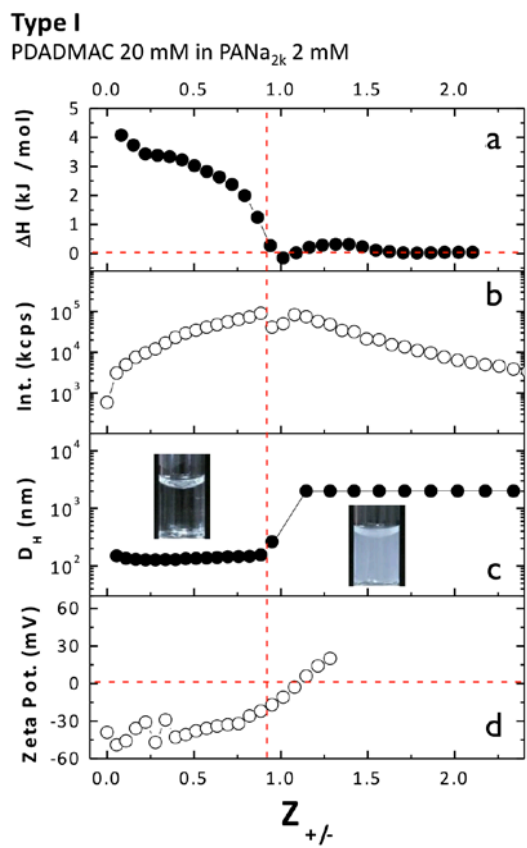

*Figure 5: Binding enthalpy, scattering intensity, hydrodynamic diameter and zeta potential observed in Type I titration. In the four experiments, PDADMAC and PANa concentrations are 20 mM and 2 mM, respectively and the same conditions of concentration, pH, temperature and injected amounts were used. At critical charge ratio 0.88 (dashed vertical line), the mixed polymer solution undergoes a liquid-liquid phase separation (coacervation).*

The presence of the PDADMAC/PANa$_{2K}$ coacervate droplets is confirmed by optical microscopy, and images of coexisting phases obtained by titration are similar to those of Fig. 2. Above the critical ratio, light scattering is not appropriate to evaluate the sizes of the structures formed, and $D_H$ was set at 2000 nm for convenience. The decrease of the scattered light at the transition (Fig. 5c) is attributed to sedimentation and/or coalescence of the coacervate phase. In parallel, zeta potential measurements evidence that the 130 nm PECS formed at low charge ratio are negatively charged, with a zeta potential of $-43\ mV$.

Electrophoretic measurements of PANa$_{2K}$ and PDADMAC 2 mM solutions (corresponding to $Z_{+/-} = 0$ and $Z_{-/+} = 0$ respectively) were performed as a control. Due to the lack of sensitivity of the Zetasizer at such low concentrations, the results were not conclusive, indicating that the data of Fig. 5d are indeed related to the PECs detected by light scattering. Upon addition of PDADMAC, $\zeta$ increases progressively, the separation occurring beyond -22 mV. With further addition of polycations, the zeta potential continues to increase and becomes slightly positive above charge neutrality.

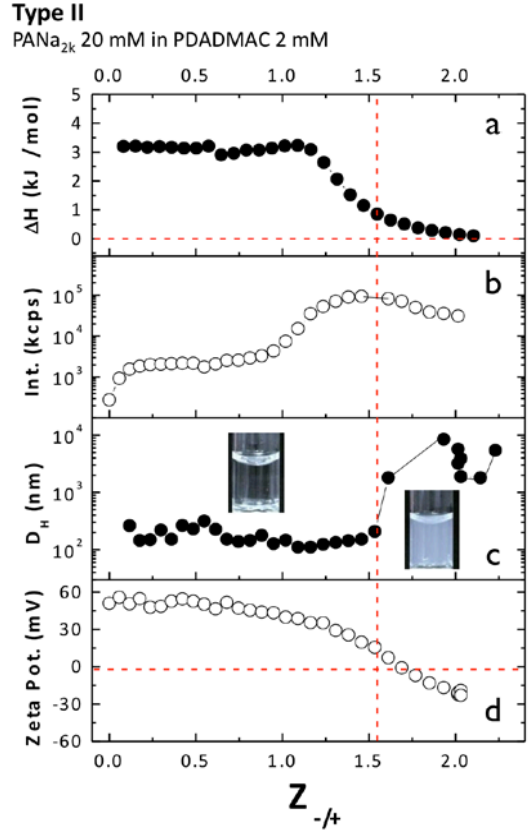

*Figure 6: Same as in Fig. 5 for Type II experiment (PANa is added to PDADMAC at respective concentration 20 and 2 mM). At the critical charge ratio 1.53 (dashed vertical line), the mixed polymer solution undergoes a liquid-liquid phase separation (coacervation).*

The data for Type II titration (Fig. 6) reveal a similar scenario: in a first regime, PECs are formed with size 140 ± 20 nm and zeta potential $+53\ mV$. This stage is followed by the coacervate phase separation at the critical ratio 1.5. The separation occurs in a range where the observed enthalpy is close to zero and the zeta potential decreased below +16 mV. Figs. 5 and 6 stress the role of the mixing order on the structures and phases formed[49,50]. The complexation observed by addition of polycations or of polyanions is qualitatively alike, but the values for the binding enthalpy, the hydrodynamic sizes and the surface charges differ (Table I). The combination of the three techniques confirms the sequential character of the titration between polyelectrolytes: with minute addition of polymer, there is the formation of stable and charged PECs. The stability of the complexes is maintained by electrostatic repulsions.[33,37,38] With further addition of opposite charged chains, more PECs form and their electrostatic charge decreases. At charge ratios close to 1, the dispersion undergoes a transition towards the coacervate phase.

|  | $D_H$ (nm) | $\zeta$ (mV) at low Z | Z of transition | $\zeta$ (mV) at transition |
|---|---|---|---|---|
| **Type I** PDADMAC in PANa | 130 ± 10 | -43 | 0.88 | -22 |
| **Type II** PANa in PDADMAC | 140 ± 20 | +53 | 1.53 | +16 |

**Table I:** *Characteristics values for hydrodynamic diameter ($D_H$) and zeta potential ($\zeta$) of PDADMAC/PANa$_{2K}$ complexes. The last two columns indicate the charge ratio and zeta potential at the coacervate transition, as found in Figs. 5 and 6.*

### 3.4 – Analysis of ITC data using the two-step MNIS model

The ITC data are analyzed using a modified version of the Multiple Non-interacting Sites (MNIS) model.[1,3,24] The MNIS model assumes that the species to be titrated, called "macromolecules" have several anchoring sites to which ligands can bind, and that the binding probability is independent on the rate of occupation of the other sites on the same macromolecule. The reaction between macromolecules and ligands is associated with an absorption or a release of heat that is proportional to the amount of binding. The reaction is characterized by a binding enthalpy $\Delta H_b$, a binding constant $K_b$ and by a reaction stoichiometry noted $n$. $n$ denotes the number of non-interacting binding sites available on each macromolecule ($n = 1$ defines the single site binding model[1]). In an ITC assay, the heat exchange is measured incrementally by successive addition of small amounts of ligands (10 µL). As a result, the measured quantity represents the derivative of the heat with respect to the ligand concentration. Introducing the charge ratio Z *in lieu* of the ligand and macromolecules concentrations, the enthalpy obtained from thermograms reads:[3,24,40]

$$\Delta H(Z, n, r) = \frac{1}{2}\Delta H_b \left(1 + \frac{n - Z - r}{\sqrt{(n + Z + r)^2 - 4Zn}}\right) \quad (2)$$

where r = 1/K$_b$[M] and [M] the molar concentration of macromolecules. Eq. 2 displays a sigmoidal decrease of the exchanged heat as a function of Z as expected from a Langmuir-type binding isotherm. Eq. 2 is equivalent to the expression provided with the microCal$^©$ software (VP-ITC).

To account for the two-step titration observed experimentally, it is assumed that the heat exchange observed during complexation is the sum of two contributions, $\Delta H_A(Z, n_A, r_A)$ for the PEC formation (the index "A" stands here for aggregates) and $\Delta H_C(Z, n_C, r_C)$ for the coacervation, both functions being of the form of Eq. 2. The binding enthalpies of each process are noted $\Delta H_b^A$ and $\Delta H_b^C$, and the stoichiometry coefficient $n_A$ and $n_C$. $r_A$ and $r_C$ are related to the binding constants $K_b^A$ and $K_b^C$ through the expressions: $r_A = 1/K_b^A[M]$ and $r_C = 1/K_b^C[M]$. Since the processes of PEC formation and coacervation have been shown to be sequential, it is assumed that the total enthalpy change during titration can be written as:

$$\Delta H(Z) = \Delta H_A(Z, n_A, r_A) + \alpha(Z)\Delta H_C(Z, n_C, r_C) \tag{3}$$

where the function $\alpha(Z)$ is the fraction of the coacervate phase at charge ratio Z. For convenience, $\alpha(Z)$ is taken of the form: $\alpha(Z) = \left(1 + exp(-(Z - Z_0)/\sigma)\right)^{-1}$, which describes a step function centered at $Z_0$ and of lateral extension $\sigma$. In essence, this model is similar to that used by Kataoka and coworkers[9,10] and more recently by Pfritis[18,19] to describe the titration of oppositely charged species in solutions. In contrast to these reports however, each step of the titration here are linked to structural changes.

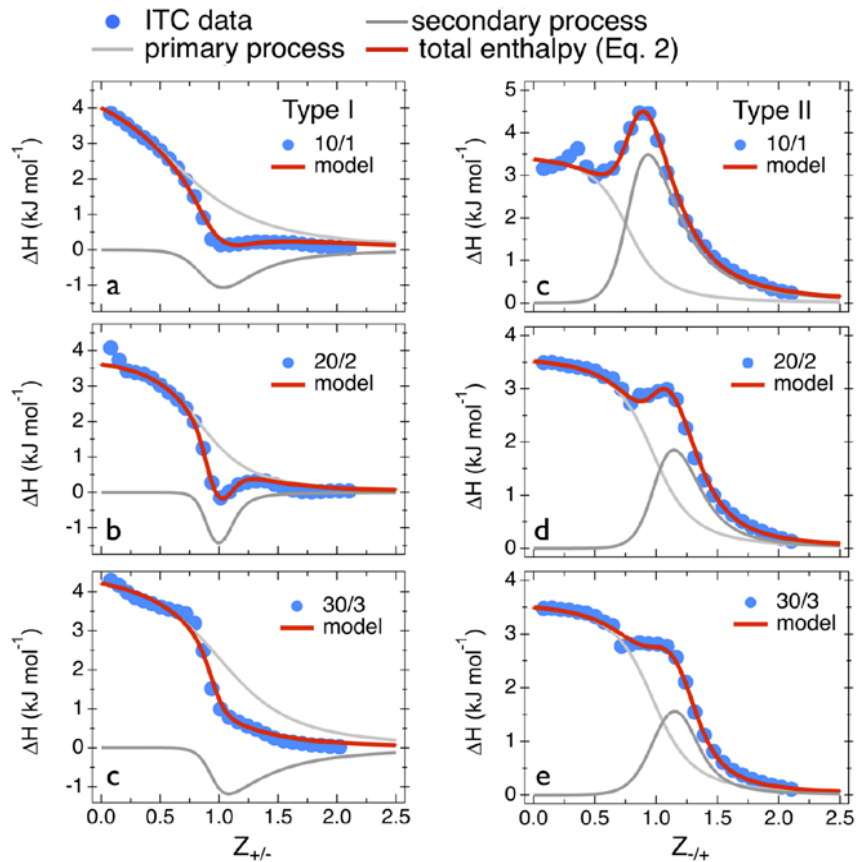

*Figure 7:* Binding isotherms for PDADMAC/PANa titrations at different concentrations and for Type I (left panels) and Type II (right panels) experiments. The continuous lines through the data points are from Eq. 3, with parameters indicated in Table II.

Pfritis *et al.* applied the concept of sequential processes too, but chose for the coacervation enthalpy a Gaussian-shape function, which is different from Eq. 3. The ITC data obtained for the titration of Type I (with $Z = Z_{+/-}$) and II (with $Z = Z_{-/+}$) are adjusted using Eq. 3, considering $\Delta H_b^A$, $\Delta H_b^C$, $n_A$, $n_C$, $r_A$ and $r_C$ as adjustable parameters. For the fitting, some constraints were imposed on the parameters. In particular we assumed that $Z_0 = n_A$ and $\sigma = 0.1$ for all curves. With the later constraint, the Z-extension of the $\alpha(Z)$-function corresponds approximately to the range where the differential power exhibits long relaxations (as in Figs. 4, S4 and S5). Fig. 7 displays the ITC data together with the fitting curves obtained thanks to Eq. 3. The separate contributions $\Delta H_A(Z)$ for PECs and $\Delta H_C(Z)$ for the coacervation are also indicated in the graphs in light and dark grey. Table II shows the set of parameters obtained for PDADMAC and PANa$_{2K}$ for Type I and Type II experiments. The results that emerge from this analysis are that *i)* the PEC formation is endothermic for both mixing orders, *ii)* the coacervation is exothermic for the addition of PDADMAC in PANa$_{2K}$ (with a $\Delta H_b^C$ = - (2.1 – 3.0) kJ mol$^{-1}$) and endothermic for the addition of PANa$_{2K}$ in PDADMAC (with a $\Delta H_b^C$ = + (2.5 – 7.0) kJ mol$^{-1}$). The entropies are positive, at + (60 – 90) J mol$^{-1}$K$^{-1}$ and + (90 – 110) J mol$^{-1}$K$^{-1}$ respectively, overcoming the unfavorable enthalpy of binding and showing that the PEC formation is entropically driven. The same results are obtained with PANa$_{100K}$, suggesting that there is little or no dependency on the molecular weight (S5). The binding constants are also very close for the two processes, around $10^4$ M$^{-1}$.

| Primary process | $\Delta H_b^A$ (kJ mol$^{-1}$) | $K_b^A$ (M$^{-1}$) | $n_A$ | $\Delta G^A$ (kJ mol$^{-1}$) | $\Delta S^A$ (J mol$^{-1}$K$^{-1}$) |
|---|---|---|---|---|---|
| **Type I PDADMAC in PANa** | | | | | |
| 10/1 | + 5.0 | 5.0 x 10$^3$ | 0.8 | - 21.1 | + 87.5 |
| 20/2 | + 3.8 | 8.3 x 10$^3$ | 0.9 | - 22.4 | + 87.9 |
| 30/3 | + 4.6 | 3.3 x 10$^3$ | 1.1 | - 20.1 | + 82.8 |
| **Type II PANa in PDADMAC** | | | | | |
| 10/1 | + 3.5 | 3.3 x 10$^4$ | 0.8 | - 25.8 | + 98.3 |
| 20/2 | + 3.4 | 1.6 x 10$^4$ | 1.0 | - 24.1 | + 92.0 |
| 30/3 | + 3.6 | 1.1 x 10$^4$ | 1.0 | - 23.1 | + 89.5 |

| Secondary process | $\Delta H_b^C$ (kJ mol$^{-1}$) | $K_b^C$ (M$^{-1}$) | $n_C$ | $\Delta G^C$ (kJ mol$^{-1}$) | $\Delta S^C$ (J mol$^{-1}$K$^{-1}$) |
|---|---|---|---|---|---|
| **Type I PDADMAC in PANa** | | | | | |
| 10/1 | - 2.6 | 2.0 x 10$^4$ | 1.1 | - 24.5 | + 73.5 |
| 20/2 | - 2.1 | 1.0 x 10$^5$ | 1.1 | - 28.5 | + 88.6 |
| 30/3 | - 3.0 | 3.3 x 10$^3$ | 1.1 | - 20.1 | + 57.3 |
| **Type II PANa in PDADMAC** | | | | | |
| 10/1 | + 7.0 | 2.5 x 10$^4$ | 1.05 | - 25.1 | + 107.6 |
| 20/2 | + 3.9 | 3.3 x 10$^4$ | 1.3 | - 25.8 | + 99.6 |
| 30/3 | + 2.5 | 3.3 x 10$^4$ | 1.3 | - 25.8 | + 94.9 |

*Table II: Thermodynamic parameters for primary (index A) and secondary (index C) processes obtained from the adjustment of the ITC curves with Eq. 3. $\Delta H_b$, $K_b$, n, $\Delta G$ and $\Delta S$ denote the binding enthalpy, binding constant, stoichiometry, free energy and entropy changes respectively.*

A second important result concerns the values of the stoichiometry coefficients $n_A$ and $n_C$. In Type I and II titrations, the values for the PEC formation are both found at 0.8 (Tab. II). In terms of positive to negative charge ratio, these values correspond to $\frac{[+]}{[-]} = 0.8$ and 1.25 respectively. Type I complexes have thus an excess of negative charges, whereas Type II complexes have an excess of positive charges. These conclusions are in good agreement with the zeta potential measurements. In contrast, the coacervation is characterized by $n_C$ coefficients close to 1 ($n_C$ = 1.1 and 1.05), suggesting that in the coacervate droplets positive and negative charges compensate. It can be concluded that the secondary ITC transition is associated with a change in stoichiometry. In one case, the transition towards the phase separation occurs with in increase of charge ratio, from $\frac{[+]}{[-]} = 0.8$ to 1.1, and in the second case with a decrease of charge ratio, from $\frac{[+]}{[-]} = 1.25$ to 0.95. This asymmetry could be at the origin of the exo/endothermic transition seen in the two titration modes. These findings point out that for strongly interacting systems the formulation pathway and mixing order matter[49,50], a result that was underestimated with regard to the ITC technique.

# IV – Conclusion

In this work, we study the complexation between oppositely charged polyelectrolytes combining isothermal titration calorimetry, light scattering and electrophoresis. Upon addition of PDADMAC to PANa or *vice-versa*, the results of the three techniques agree well with each other, and reveal the succession of a two-step titration. The primary process is the formation of highly charged complexes. Their sizes are around 100 nm, and their charges are those of the dominant polyelectrolyte, negative for the titration of PDADMAC in PANa$_{2K}$ and positive in the reverse case. It is suggested that the stability of the complexes arises from electrostatic repulsions.[33,37,38] The second process occurs around charge stoichiometry and is related to the transition from the PECs towards coacervate droplets. As for the titration calorimetry response, the phase separation displays an exothermic profile upon addition of PDADMAC in PANa$_{2K}$, and an endothermic profile for the reverse. A schematic representation of the various phases and transitions induced by titration can be found in Fig. 8. A useful perspective is gained by analyzing the ITC data quantitatively thanks to a modified version of the Multiple Non-interacting Sites model.[3,24,40] The model assumes that the coacervation is kinetically activated and starts only after the formation of the PECs. In the transition range, the titration calorimeter displays long transients associated with the formation of the coacervate droplets.[24] The results that emerge from the analysis is that small positive enthalpies and large positive entropies are found uniformly for both complex formation and coacervation, in agreement with earlier results.[5,6,10,16-19,24] Once translated in terms of thermal energy per titrating charges, the enthalpy costs of the reactions correspond to $+1k_BT$, whereas the gains in entropy are of the order of $+10k_BT$. At this point, it is not clear whether the enthalpies for the first process is related to the pairing of the opposite charges only,[19] or to some collective phenomena involving the PEC formation. This issue needs to be investigated further.

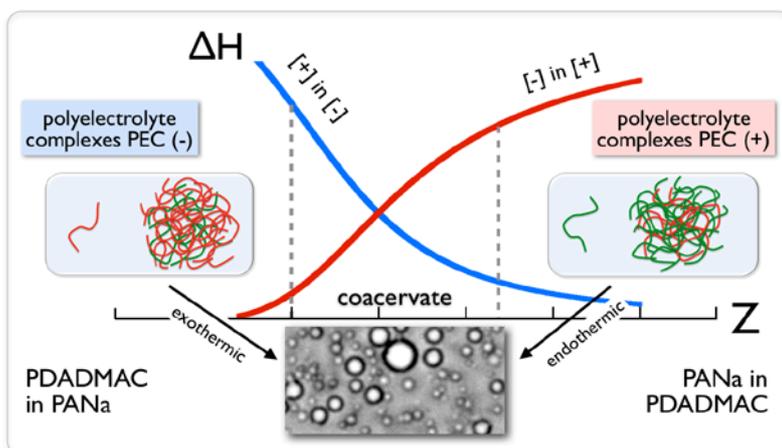

*Figure 8:* Schematic representation of the thermodynamic responses and structural changes obtained by titrating oppositely charged polymers.

As mentioned in the introduction, ITC is probably the sole simple technique to retrieve the thermodynamic parameters of a physic-chemical reaction, where changes in free energy are of the order of 1-10 $k_B T$ per titrating ligand. Our work on PDADMAC and PANa complexation emphasizes however that it is a very delicate technique and that binding isotherms alone are in general not sufficient to identify the key mechanisms involved in titration. To achieve this goal here, we had to survey the impact of concentration, pH, molecular weight and mixing order. Only after summing up the different results (and helped by structural and microscopy data), were we able to conclude about the nature of the transitions and propose an appropriate modeling. The above reasons may explain the differences between our findings and those of Alonso and coworkers.[17] In conclusion, the present results confirm the existence a general two-step behavior in thermodynamic titration of oppositely charged poly(diallyldimethylammonium chloride) and poly(sodium acrylate). The knowledge gained from synthetic polymers will benefit the formulations of oppositely charged systems for the fabrication of electrostatic based structures and materials.

## Supporting Information

The Supporting Information includes sections on the acid-base titration of poly(acrylic acid)/poly(sodium acrylate) (S1), images of dispersions prepared by direct mixing at pH7 (S2), evidence of pH change for titrations at pH7 (S3), ITC experiments and modeling at pH7 (S4) and using PANa$_{100K}$ (S5). All authors have given approval to the final version of the manuscript.

# Acknowledgment

We thank Frédéric Loosli, Fanny Mousseau, Emek Seyrek and Serge Stoll for fruitful discussions. This work was supported by a bilateral cooperation between CNRS and FAPESP (Proc. No. 2010/52411-6). L.V. thanks the Brazilian Agency Capes for a PhD fellowship and the CNPq (Conselho Nacional de Desenvolvimento Científico e Tecnológico) in Brazil for postdoctoral fellowship. ANR (Agence Nationale de la Recherche) and CGI (Commissariat à l'Investissement d'Avenir) are gratefully acknowledged for their financial support of this work through Labex SEAM (Science and Engineering for Advanced Materials and devices) ANR 11 LABX 086, ANR 11 IDEX 05 02. This research was supported in part by the Agence Nationale de la Recherche under the contract ANR-13-BS08-0015 (PANORAMA).

# Supporting Information

**Evidence of a two-step process and pathway dependency in the thermodynamics of poly(diallyldimethylammonium chloride)/ poly(sodium acrylate) complexation**

L. Vitorazi[1,2], N. Ould-Moussa[1], S. Sekar[3], J. Fresnais[4], W. Loh[2], J.-P. Chapel[3] and J.-F. Berret[1]

[1]*Matière et Systèmes Complexes, UMR 7057 CNRS Université Denis Diderot Paris-VII, Bâtiment Condorcet, 10 rue Alice Domon et Léonie Duquet, 75205 Paris, France.*
[2]*Institute of chemistry, Unversidade Estadual de Campinas (UNICAMP), Caixa Postal 6154, Campinas, São Paulo, Brazil.*
[3]*Centre de Recherche Paul Pascal (CRPP), UPR CNRS 8641, Université Bordeaux 1, 33600 Pessac, France*
[4]*Physicochimie des Electrolytes et Nanosystèmes interfaciaux (PHENIX) UMR 7195 CNRS-UPMC, 4 place Jussieu, 75252 Paris, France*

**S1 – Determination of degree of ionization of poly(acrylic acid)/poly(sodium acrylate) as a function of pH**
**S2 – Images of dispersions prepared by direct mixing at pH7**
**S3 - Evidence of pH changes during titrations performed at pH7**
**S4 - ITC experiments between PDADMAC and PANa$_{2K}$ at pH7**
**S5 – ITC experiments between PDADMAC and PANa$_{100K}$**

**S1 - Determination of degree of ionization of poly(acrylic acid)/poly(sodium acrylate) as a function of pH**

Fig. S1 shows the acido-basic titration of poly(sodium acrylate) ($M_w$ = 2100 g mol$^{-1}$) by addition of sodium hydroxide solution (NaOH). The continuous line corresponds to the derivative of the pH as a function of the number of added moles, dpH/dn$_{NaOH}$. In this experiment, 0.1284 g of PANa, corresponding to 1.36 10$^{-3}$ mole of carboxylic groups was titrated with 7.388 mL (distance between the two maxima) of NaOH prepared at the concentration of 0.133 mol L$^{-1}$. From the titration, we found that the amount of carboxylic groups represent 72% of the monomers. There is thus less carboxylic acid monomers as expected from the calculation. This discrepancy could originate, in part from the presence of bound water molecules in the polymer powder. The percentage of 72% was used in the estimation of the anionic charge coming from the PANa chains.

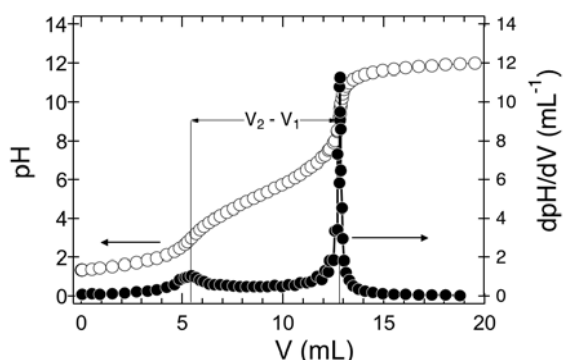

***Figure S1****: Potentiometric curves for the increment addition of NaOH to poly(acrylic acid) of molecular weight 2100 g mol$^{-1}$.*

## S2 – Images of dispersions prepared by direct mixing at pH7

Fig. S2 displays images of PDADMAC/ PANa$_{2K}$ mixed solutions obtained at pH7. The solutions were obtained by direct mixing protocols. Turbid samples are associated with a liquid-liquid phase separation (coacervation).

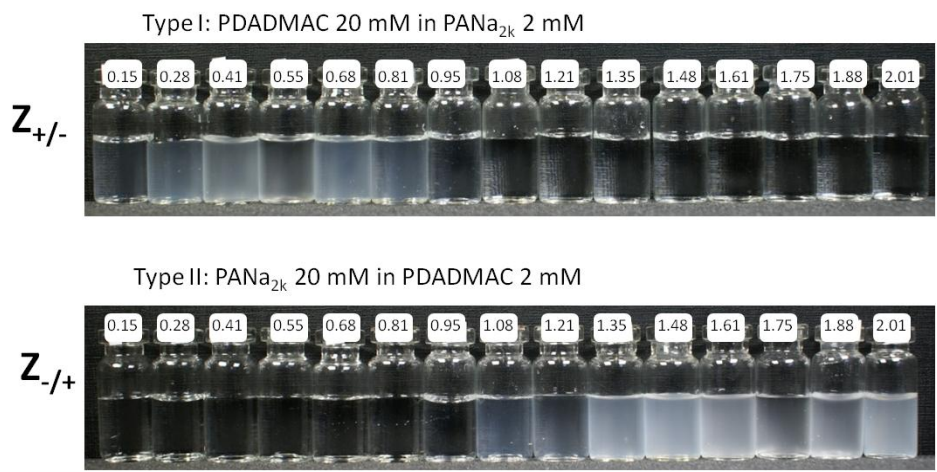

*Figure S2:* Direct mixtures of PANa$_{2K}$ and PDADMAC for Type I and II experiments at pH 7.

## S3 - Evidence of pH changes during titrations performed at pH7

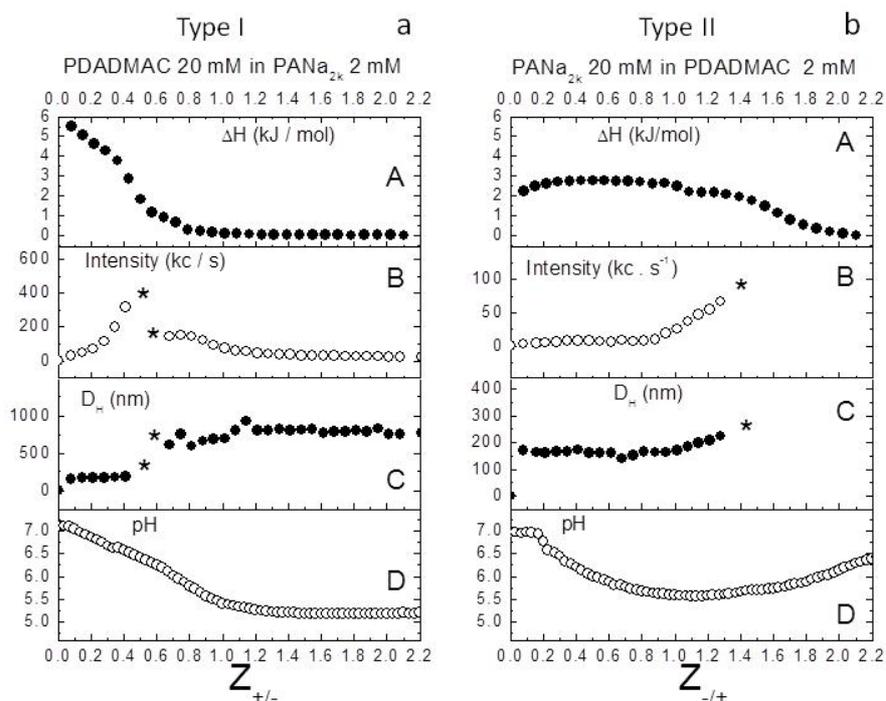

*Figure S3*: Binding enthalpy (A), light scattered intensity (B), hydrodynamic diameter (C) and pH measurements (D) found by titration of PANa$_{2K}$ by PDADMAC (Type I experiment) and of PDADMAC by PANa$_{2K}$ (Type II experiment). Prior to titration, the pH of the initial polymer solutions was set at pH 7.

## S4 - ITC experiments between PDADMAC and PANa₂K at pH7

Figs. S4 show the thermograms (a,d) and binding isotherms with (b,e) for Type I and II titrations between PDADMAC in PANa$_{2K}$ at *pH*7. The characteristic features of the ITC curves are identical to those observed with PANa$_{2K}$ at pH10. Upon addition of PANa$_{2K}$ to PDADMAC or *vice-versa*, ITC reveals the existence of two sequential processes, one endothermic at low charge ratio, and the second being either exo- or endothermic depending on the mixing order.

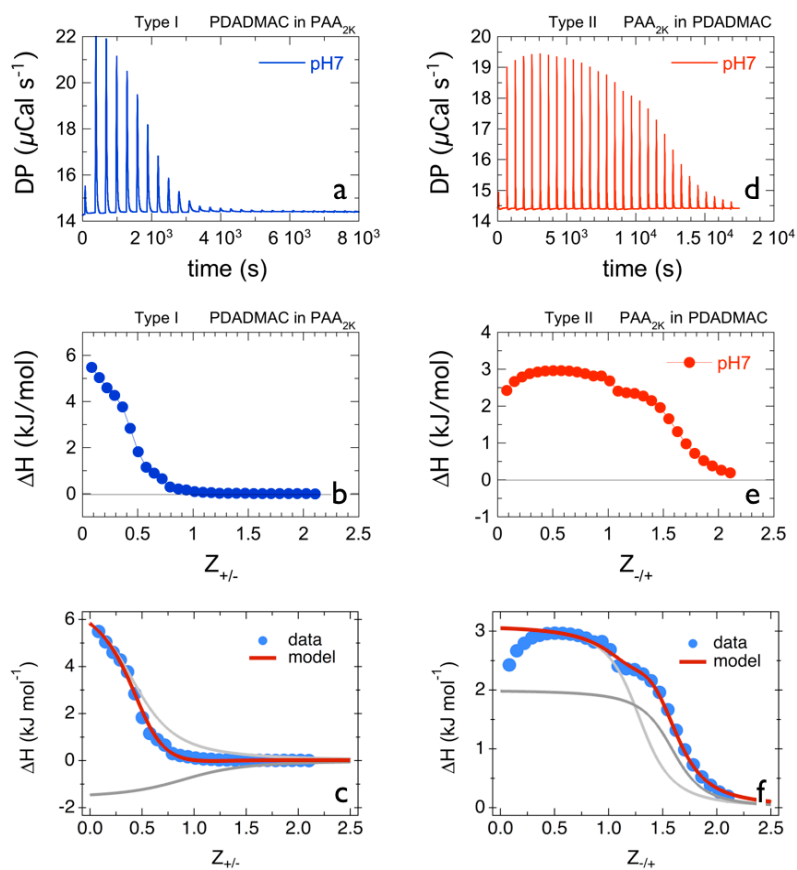

***Figure S4***: *ITC curves for addition of PDADAMAC in PANa$_{2K}$ (a,b,c) and PANa$_{2K}$ in PDADMAC (d,e,f) at pH7. In c) and f) the binding enthalpy curves are adjusted using the model described in the main text (Eq. 3).*

In Figs. S4c and S4f are plotted the binding isotherms together with the adjustments using Eq. 3. The different parameters retrieved from the adjustment are listed in Table S4. The legends are the same as those of Fig. 7. The present approach shows that at pH7 and pH10, the thermodynamics of titration remains the same, and that the molecular weight of the polymer does not play a major role in the sequence of reactions.

| Primary process | $\Delta H_b^A$ (kJ mol$^{-1}$) | $K_b^A$ (M$^{-1}$) | $n_A$ | $\Delta G^A$ (kJ mol$^{-1}$) | $\Delta S^A$ (J mol$^{-1}$K$^{-1}$) |
|---|---|---|---|---|---|
| **Type I** PDADMAC in PANa$_{2K}$ | | | | | |
| pH7 (20/2) | + 7.0 | 5.0 x 10³ | 0.5 | - 21.0 | +94.3 |
| Type II PANa$_{2K}$ in PDADMAC pH7 (20/2) | + 3.1 | 2.5 x 10$^4$ | 1.3 | - 25.1 | + 94.6 |
| | | | | | |
| Secondary process | $\Delta H_b^C$ (kJ mol$^{-1}$) | $K_b^C$ (M$^{-1}$) | $n_C$ | $\Delta G^C$ (kJ mol$^{-1}$) | $\Delta S^C$ (J mol$^{-1}$K$^{-1}$) |
| Type I PDADMAC in PANa$_{2K}$ pH7 (20/2) | - 1.6 | 5.0 x 10$^3$ | 1.0 | - 21.1 | + 65.4 |
| Type II PANa$_{2K}$ in PDADMAC pH7 (20/2) | + 2.0 | 3.3 x 10$^4$ | 1.6 | - 25.8 | + 93.2 |

*Table S4:* *List of the thermodynamic parameters determined for the binding enthalpies between PDADMAC and PANa$_{2K}$ at pH7.*

### S5 - ITC experiments between PDADMAC and PANa$_{100K}$

Figs. S5 show the thermograms (a,d) and binding isotherms with (b,e) for Type I and II titrations between PDADMAC in PANa$_{100K}$. The ITC data were obtained at *pH* 10 and T = 25 °C. The characteristic features of the ITC curves are identical to those observed with PANa$_{2K}$. Upon addition of PANa$_{100K}$ to PDADMAC or *vice-versa*, ITC reveals the existence of two sequential reactions, one endothermic at low charge ratio, and the second being either exo- or endothermic depending on the mixing order.

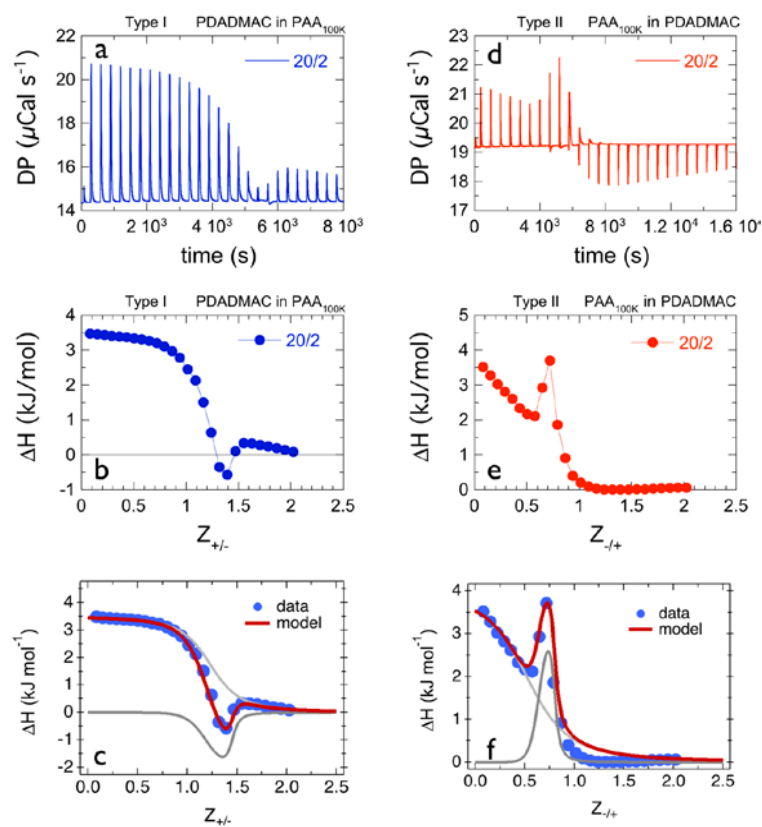

***Figure S5****: ITC curve for addition of PDADAMAC in PANa$_{100K}$ at pH 10 (a,b,c) and PANa$_{100K}$ in PDADMAC (d,e,f). In c) and f) the binding enthalpy curves are adjusted using the model described in the main text (Eq. 2).*

In Figs. S5c and S5f are plotted the binding isotherms together with the adjustments using Eq. 3. The different parameters retrieved from the adjustment are listed in Table S5. The legends are the same as those of Fig. 7. The present approach shows that for PANa$_{2K}$ and PANa$_{100K}$, the thermodynamics of titration remains the same, and that the molecular weight of the polymer does not play a major role in the sequence of reactions.

| **Primary process** | $\Delta H_b^A$ (kJ mol$^{-1}$) | $K_b^A$ (M$^{-1}$) | $n_A$ | $\Delta G^A$ (kJ mol$^{-1}$) | $\Delta S^A$ (J mol$^{-1}$K$^{-1}$) |
|---|---|---|---|---|---|
| **Type I** PDADMAC in PANa$_{100K}$ 20/2 | + 3.5 | 2.5 x 10$^4$ | 1.25 | - 25.1 | + 95.9 |
| **Type II** PANa$_{100K}$ in PDADMAC 20/2 | + 4.0 | 6.3 x 10$^3$ | 0.6 | - 21.6 | + 86.1 |
| **Secondary process** | $\Delta H_b^C$ (kJ mol$^{-1}$) | $K_b^C$ (M$^{-1}$) | $n_C$ | $\Delta G^C$ (kJ mol$^{-1}$) | $\Delta S^C$ (J mol$^{-1}$K$^{-1}$) |
| **Type I** PDADMAC in PANa$_{100K}$ 20/2 | - 2.2 | 5.0x 10$^5$ | 1.45 | - 32.5 | + 101.7 |
| **Type II** PANa$_{100K}$ in PDADMAC 20/2 | + 3.5 | 5.0 x 10$^5$ | 0.8 | - 32.5 | + 120.8 |

***Table S5:*** *List of the thermodynamic parameters determined for the binding enthalpies between PDADMAC and PANa$_{100K}$ at pH10.*